
\input epsf

\ifx\epsffile\undefined\message{(FIGURES WILL BE IGNORED)}
\def\insertplot#1#2{}
\def\insertfig#1#2{}
\else\message{(FIGURES WILL BE INCLUDED)}
\def\insertplot#1#2{
\midinsert\centerline{{#1}}\vskip0.2truein\centerline{{\epsfxsize=\hsi
ze
\epsffile{#2}}}\vskip0.5truecm\endinsert}
\def\insertfig#1#2{
\midinsert\centerline{{\epsfxsize=\hsize\epsffile{#2}}}\vskip0.2truein

\centerline{{#1}}\vskip0.5truecm\endinsert}
\fi

\input harvmac

%
%
%
%
%
\ifx\answ\bigans
\else
\output={

\almostshipout{\leftline{\vbox{\pagebody\makefootline}}}\advancepageno

}
\fi
%
%
%

%
%

%
%
\def\UCSD#1#2{\noindent#1\hfill #2%
\bigskip\supereject\global\hsize=\hsbody%
\footline={\hss\tenrm\folio\hss}}
%
%
\def\abstract#1{\centerline{\bf Abstract}\nobreak\medskip\nobreak\par
#1}
%
%
%
%
\edef\tfontsize{ scaled\magstep3}
 \tfontsize  \tfontsize
\font\titlermss=cmr5 \tfontsize \font\titlei=cmmi10 \tfontsize
\font\titleis=cmmi7 \tfontsize \font\titleiss=cmmi5 \tfontsize
\font\titlesy=cmsy10 \tfontsize \font\titlesys=cmsy7 \tfontsize
\font\titlesyss=cmsy5 \tfontsize  \tfontsize
\skewchar\titlei='177 \skewchar\titleis='177 \skewchar\titleiss='177
\skewchar\titlesy='60 \skewchar\titlesys='60 \skewchar\titlesyss='60
\scriptscriptfont0=\titlermss
\scriptscriptfont1=\titleiss
\scriptscriptfont2=\titlesyss
%
%
%

%
\def\inv{^{\raise.15ex\hbox{${\scriptscriptstyle -}$}\kern-.05em 1}}
\def\lbar{{\lower.35ex\hbox{$\mathchar'26$}\mkern-10mu\lambda}}

%
%
%
%
\def\dsl{\,\raise.15ex\hbox{/}\mkern-13.5mu D} 
\def\delsl{\raise.15ex\hbox{/}\kern-.57em\partial}
\def\Ksl{\hbox{/\kern-.6000em\rm K}}
\def\Asl{\hbox{/\kern-.6500em \rm A}}
\def\Dsl{\hbox{/\kern-.6000em\rm D}} 
\def\Qsl{\hbox{/\kern-.6000em\rm Q}}
\def\gradsl{\hbox{/\kern-.6500em$\nabla$}}
%
%
\def\lspace{\ifx\answ\bigans{}\else\qquad\fi}
\def\lbspace{\ifx\answ\bigans{}\else\hskip-.2in\fi} 
%
%
\def\boxeqn#1{\vcenter{\vbox{\hrule\hbox{\vrule\kern3pt\vbox{\kern3pt
        \hbox{${\displaystyle #1}$}\kern3pt}\kern3pt\vrule}\hrule}}}
%
%
\def\mbox#1#2{\vcenter{\hrule \hbox{\vrule height#2in
\kern#1in \vrule} \hrule}}
%
%
%
%

%
%
%
%
%

%

\def\bar#1{\overline{#1}}

\def\ket#1{\left| #1\right\rangle}

\def\darr#1{\raise1.5ex\hbox{$\leftrightarrow$}\mkern-16.5mu #1}

%
%
\def\frac#1#2{{\textstyle{#1\over #2}}} 
%
%
%
%

\def\GeV{{\rm GeV}}

%
%
%
%
%
%
\def\ltap{\ \raise.3ex\hbox{$<$\kern-.75em\lower1ex\hbox{$\sim$}}\ }
\def\gtap{\ \raise.3ex\hbox{$>$\kern-.75em\lower1ex\hbox{$\sim$}}\ }
\def\gl{\ \raise.5ex\hbox{$>$}\kern-.8em\lower.5ex\hbox{$<$}\ }
\def\roughly#1{\raise.3ex\hbox{$#1$\kern-.75em\lower1ex\hbox{$\sim$}}}
%
%

%

%

\def\pl#1#2#3{{Phys. Lett. } {#1}B (#2) #3}

\def\physrev#1#2#3{{Phys. Rev. } {#1} (#2) #3}

\relax

\def\({\left(}
\def\){\right)}
\def\cbar{\bar c}
\def\bbar{\bar b}
\def\qbar{\bar q}
\def\diq{{(cc)}}
\def\bc{{(bc)}}
\def\3bar{\overline{3}}
\def\had{{\Lambda_{\rm QCD}}}

\noblackbox
 at 12truept
\vskip 1.in
\centerline{{\titlefont{Heavy Quark Fragmentation to}}}
\medskip
\centerline{{\titlefont{Baryons Containing Two Heavy Quarks}}}
\bigskip
\centerline{Adam F.~Falk{}$^{a}$
\footnote{$^\dagger$}{Address after 10/1/93: Dept. of Physics, UC San
Diego, La Jolla, CA 92093},
Michael Luke{}$^{b}$ \footnote{$^\ddagger$}{Address after 9/1/93: Dept.
of Physics, University of Toronto, Toronto, Ontario, Canada M5S 1A7},
Martin J.\ Savage{}$^{b}$ \footnote{$^{\S}$}{Address after 9/1/93:
Dept. of Physics, Carnegie Mellon University, Pittsburgh, PA 15213}
\footnote{$^{\star}$} {SSC Fellow}
and Mark B. Wise{}$^{c}$}
\bigskip
\centerline{\sl a) Stanford Linear Accelerator Center, Stanford CA
94309}
\centerline{\sl b) Department of Physics, University of California at
San Diego,}\centerline{\sl 9500 Gilman Drive, La Jolla CA 92093}
\centerline{\sl c) California Institute of Technology, Pasadena CA
91125}
\vfill
\abstract{We discuss the fragmentation of a heavy quark to a baryon
containing two heavy quarks of mass $m_Q\gg\Lambda_{\rm QCD}$.
In this limit
the heavy quarks first combine perturbatively into a compact diquark
with a radius small compared to $1/\Lambda_{\rm QCD}$, which interacts
with the light hadronic degrees of freedom exactly as does a heavy
antiquark.  The subsequent evolution of this $QQ$ diquark to a $QQq$
baryon is identical to the fragmentation of a heavy antiquark to a
meson. We apply this analysis to the production of baryons of the form
$ccq$, $bbq$, and $bcq$.}
\vfill
\UCSD{\vbox{
\hbox{UCSD/PTH 93-11, CALT-68-1868, SLAC-PUB-6226}
\hbox{hep-ph/9305315}}}{May 1993}
\eject

The spectroscopy and interactions of baryons consisting of two heavy
quarks and one light quark simplify in the limit that the heavy quark
masses $m_Q$ tend to infinity.  This is because the heavy quarks are
bound into a diquark whose radius $r_{QQ}$ is much
smaller than the typical length scale $1/\had$ of nonperturbative QCD
interactions.  In the limit $r_{QQ}\ll1/\had$ the heavy diquark has
interactions with the light quark and other light degrees of freedom
which are identical to those of a heavy antiquark. Hence as far as
these light degrees of freedom are concerned, the diquark is nothing
more than the pointlike, static, colour antitriplet source of the
confining colour field in which they are bound \ref\swa
{M.J. Savage and M.B. Wise, \pl{248}{1990}{151}.}--%
\nref\wsa{M.J. White and M.J. Savage, \pl{271}{1991}{410}.}%
\ref\fsr{For related work in the context of potential models, see
S.~Fleck, B.~Silvestre-Brac and J.~M.~Richard, \physrev{D38}
{1988}{1519}.}.

One immediate result of this limit is that the spectrum of such
``doubly heavy'' baryons is related to the spectrum of mesons
containing a single heavy antiquark\swa.  It also follows that the form
factors describing their semileptonic decays may be related to the
Isgur-Wise function, which arises in the semileptonic decay of heavy
mesons \wsa.  In this note we will apply the same symmetries to the
nonperturbative dynamics which governs the production of such states
via fragmentation processes. We will use our results to estimate the
production rates for baryons of the form $ccq$, $bbq$ and $bcq$;
however, we note that, especially in the $cc$ system, the heavy
diquarks are not particularly small relative to $1/\had$, so there may
well be sizeable corrections to our results.

The fragmentation of a heavy quark $Q$ into a $QQq$ (or $QQ'q$) baryon
factorises into short-distance and long-distance contributions.  The
heavy quark first fragments into a heavy diquark via a process which is
perturbatively calculable.  In fact, the amplitude may be trivially
related to that for the fragmentation of $Q$ into quarkonium $Q\bar Q$.
The subsequent fragmentation of the diquark $QQ$ to a baryon is
identical to the fragmentation of a $\bar Q$ to a meson $\bar Qq$; this
information may be obtained from experimental data on production of
heavy mesons.

We will begin with the case of baryons of the form $QQq$, in which the
two heavy quarks have the same flavour.  For concreteness we will
discuss the production of baryons with two charm quarks $ccq$; the
extension to bottom baryons $bbq$ will be trivial.  As the colour
wavefunction of the charm quarks is antisymmetric and the quarks are
taken to be in the ground state $S$-wave, the spin wavefunction must be
symmetric.  Hence the $cc$ can only form a spin-1 diquark, which we
shall denote by $\diq$.  The $\diq$ can then fragment either to a
spin-$\frac12$ baryon, which we shall denote by $\Sigma_{cc}$, or to a
spin-$\frac32$ baryon, which we shall call $\Sigma_{cc}^*$.

The initial short distance fragmentation process $c\to\diq\cbar$ is
analogous to that for the fragmentation into charmonium, $c\to\psi c$,
which has been shown by Braaten, Cheung and Yuan \ref\ericb{E. Braaten,
K. Cheung and T.C. Yuan, NUHEP-TH-93-2,UCD-93-1 (1993)} to be
calculable in QCD perturbation theory.  The Feynman diagrams
responsible for $\diq$ production are shown in \fig\feyn{Feynman
diagrams responsible for the fragmentation $c\rightarrow\diq$.}.  The
computation, which we shall not repeat, follows directly that outlined
for $c\to\psi c$ in \ericb.  In fact, after some rearrangement the form
of the fragmentation function $D_{c\rightarrow\diq}(z)$ is exactly the
same as that for $D_{c\rightarrow\psi}(z)$, except for an overall
normalisation factor.  Since the charm quarks in the $\diq$ are in an
overall colour $\bar 3$ rather than a singlet, there is a colour factor
of $2/3$ instead of $4/3$ in the amplitude. The colour wavefunction of
the diquark is $\ket{\diq_a}=\frac12 \epsilon_{abc}\ket{c^b}\ket{c^c}$
where $a,b,c$ are colour indices.  The factor of $1/2$ in the
normalisation of the diquark state is canceled by a factor of $2$ in
the matrix element from fermi statistics, giving an overall $2/3$ in
the amplitude.  Squaring and summing over final colours then gives the
desired fragmentation function, renormalised at the scale $\mu=3m_c$:
\eqn\ctodiq{
    D_{c\to\diq}(z,\mu=3m_c) = {4\over9\pi}{ |R_\diq(0)|^2
    \over m_c^3}\alpha_s(3m_c)^2F(z)\,,}
where
\eqn\fofz{
    F(z)={z(1-z)^2\over(2-z)^6}\left(16-32z+72z^2-32z^3+5z^4\right)\,.}
Here $R_\diq(0)$ is the nonrelativistic radial wavefunction at the
origin for the perturbatively bound diquark.  Unlike the case of
charmonium, there is no physical ``decay constant'' $f_\diq$ to which
it may be related.  However we may na\"\i vely scale $R_\diq(0)$ from
charmonium by noting that in a hydrogen-like potential $R(0)$ is
proportional to $(C_F\alpha_s)^{3/2}$ where the colour factor $C_F$ is
$\frac43$ for $c\cbar$ and $\frac23$ for $cc$.  Hence we expect that
$|R_\diq(0)|^2 \approx |R_\psi(0)|^2/8\approx(0.41\;\GeV)^3$.

The complete fragmentation function $D_{c\to\Sigma}(z)$ for
$c\to\diq\cbar\to(\Sigma_{cc},\Sigma_{cc}^*)\cbar$ is given by
convolving the function $D_{c\to\diq}(z)$ with the amplitude for the
heavy antitriplet diquark to fragment to a ground state baryon with a
single light quark.  This latter function, which we shall denote
$D_{\bar Q\to M}(z)$, may be determined by data on charm and bottom
antiquark fragmentation to heavy mesons.  Summing over the
$\Sigma_{cc}$ and $\Sigma_{cc}^*$ baryons, we then find
\eqn\fragdist{
    D_{c\to\Sigma}(z) = \int_z^1 {dy\over y}
    D_{c\to\diq}(z/y)D_{\bar Q\to M}(y)\,.}
However, in the limit $m_c\gg\had$ the heavy diquark carries all of the
momentum of the baryon, and $D_{\bar Q\to M}(y) =\delta(1-y)P_{\bar
Q\to M}$, where $P_{\bar Q\to M}$ is the integrated probability for a
heavy antiquark to fragment to a ground state meson.  Then \fragdist\
takes the simpler form
\eqn\fragtot{
    D_{c\to\Sigma}(z) = P_{\bar Q\to M} D_{c\to\diq}(z)\,.}
Since in the limit we are considering all excited states will decay
strongly to the ground state, we may replace this probability by unity.
(There is a small correction from the fragmentation probability to
baryons, $P_{\bar Q\to\bar\Lambda,\bar\Sigma}$; here this would lead to
an exotic $cc\qbar\qbar$ final state.)  Then the integral of \ctodiq\
yields the final result
\eqn\fragint{
    \int_0^1 dz\;D_{c\to\Sigma}(z)={4\over9\pi}\alpha_s(3m_c)^2
    {|R_\diq(0)|^2\over m_c^3}\left({1189\over30}-57\ln 2\right)\,.}

The $\Sigma_{cc}^*$ and $\Sigma_{cc}$ baryons will be produced in the
ratio $2:1$. However we may use the polarised fragmentation functions
$D_{c\to\diq}(z)$ derived in \ref\falka{A.F. Falk, M. Luke, M.J. Savage
and M.B. Wise, UCSD/PTH 93-06, CALT-68-1864, SLAC-PUB-6175,
hep-ph/9305260 (1993)} to compute individually the populations of the
various helicity states of the $\Sigma_{cc}$ and $\Sigma_{cc}^*$. Let
$P$ be the net polarisation of the initial charm quarks.  Then the
populations of the baryon helicity states can be determined from $P$
and the fraction $\zeta$ of the produced diquarks which are
transversely rather than longitudinally aligned. For example, a charm
quark with energy $E$ and helicity $\frac12$ can fragment to a $\diq$
diquark with helicity $0$ or $1$, but not to one with helicity $-1$, to
leading order in $m_c/E$.  Hence the net polarisation of the diquarks
is degraded to $\zeta P$, where $\zeta=0.69$ \falka. To make the
baryon, the diquark must then be combined with a light quark from the
nonperturbative part of the fragmentation.  In the limit $m_c\gg\had$,
the helicity of the diquark is irrelevant to this soft process; the
parity invariance of QCD then requires than the light quarks populate
equally the helicities $\pm\frac12$.  Since the $\Sigma_{cc}$ and
$\Sigma_{cc}^*$ are produced incoherently, we may compute independently
the probabilities that the light quark and the diquark will combine
into the two possible angular momentum states, $\frac12$ and $\frac32$.
We then find the various helicity states to be populated in the ratios
$\Sigma_{cc}^*(\pm\frac32):\Sigma_{cc}^*(\pm\frac12):
\Sigma_{cc}(\pm\frac12)=\frac14\zeta(1\pm
P):\frac13+\frac14\zeta(-1\pm\frac13P):\frac16(1\pm\zeta P)$.

The quark fragmentation function \fragdist\ has been computed at the
renormalisation scale $\mu=3m_c$.  The Altarelli-Parisi equations must
then be used to evolve it up to a high scale $\mu=M$ typical of
collider energies, a procedure which sums large logarithms of the form
$\ln(m_c/M)$.  This evolution has two effects.  First, it softens the
$z$ distributions, but because the relevant splitting functions
$P_{c\to cg}(z)$ and $P_{c\to gc}(z)$ integrate to zero, the
fragmentation probability $\int_0^1 dz\,D_{c\to\Sigma}(z)$ remains
unchanged.  Second, a gluon fragmentation function
$D_{g\to\Sigma}(z,M)$ is induced via the gluon splitting function
$P_{g\to c\cbar}(z)$.  In leading logarithmic approximation, this
effect, of order $\alpha_s^3\ln(M/m_c)$, dominates over any direct
contribution to gluon fragmentation, which would be of order
$\alpha_s^3$ without the $\ln(M/m_c)$ enhancement.

We now turn to the production of baryons of the form $bcq$, in which
the heavy quarks are not identical.  In this case the $bc$ diquark may
be in either a spin-$0$ state, which we shall denote $\bc$, or a
spin-$1$ state, which we shall denote $\bc^*$.  In addition to the
formation of $\Sigma_{bc}$ and $\Sigma_{bc}^*$ baryons from the
fragmentation of the of the $\bc^*$ diquark, we now have the
possibility of the formation of $\Lambda_{bc}$ baryons from the
fragmentation of the $\bc$ diquark.

Just as in the case of diquarks consisting of two identical heavy
quarks, it is straightforward to relate the fragmentation functions for
$b\to\bc^{(*)}\cbar$ and $c\to\bc^{(*)}\bbar$ to those for production
of the physical states $B_c$ and $B_c^*$.  These latter functions have
been calculated by Braaten, Cheung and Yuan \ref\ericc{E. Braaten, K.
Cheung and T.C. Yuan, NUHEP-TH-93-6, UCD-93-9 (1993)}, and we use their
computation to obtain our results.  Unlike the $\diq$ case, the colour
wavefunction of the diquark is $\ket{\bc_a}=\frac{1}{\sqrt{2}}
\epsilon_{abc}\ket{b^b}\ket{c^c}$, and there is no factor of 2 from
fermi statistics, so the spin-$0$ diquark fragmentation function with
an initial bottom quark is given by
\eqn\btobc{
    D_{b\to\bc}(z,\mu_0) = {2\over9\pi}{|R_\bc(0)|^2
    \over m_c^3}\alpha_s(\mu_0)^2F(z,r)\,,}
where
\eqn\fofzr{\eqalign{
    F(z,r)={rz(1-z)^2\over12(1-(1-r)z)^6}\big[&6-18(1-2r)z
    +(21-74r+68r^2)z^2\cr -2(1-r)(&6-19r+18r^2)z^3
    +3(1-r)^2(1-2r+2r^2)z^4\big]\,,\cr}}
$r=m_c/(m_c+m_b)$, and $\mu_0=\sqrt{4m_c(m_b+m_c)}$ is a mass scale
intermediate between $m_c$ and $m_b$.  (The selection of this scale is
discussed in detail in \ericc.)  The analogous function for an initial
charm quark, $D_{c\to\bc}(z)$, is given by \btobc\ with the
replacements $F(z,r)\to(m_c/m_b)^3 F(z,1-r)$ and
$\mu_0\to\mu_0'=\sqrt{4m_b(m_b+m_c)}$.  Finally, the fragmentation
function $D_{b\to\bc^*}(z)$ to the spin-$1$ diquark is given by \btobc\
with $F(z,r)\to F^*(z,r)$, where
\eqn\fstar{\eqalign{
    F^*(z,r)={rz(1-z)^2\over4(1-(1-r)z)^6}\big[&2-2(3-2r)z
    +3(3-2r+4r^2)z^2\cr -2(1-&r)(4-r+2r^2)z^3
    +(1-r)^2(3-2r+2r^2)z^4\big]\,.\cr}}
Then the same replacements $F^*(z,r)\to(m_c/m_b)^3 F^*(z,1-r)$ and
$\mu_0\to\mu_0'$ yield $D_{c\to\bc^*}(z)$.

A calculation similar to that in Ref.~\falka\ gives $\zeta=0.69$ for
the net alignment of the $\bc^*$.  Hence the populations of the various
helicity states of the $\Sigma_{bc}$ and $\Sigma_{bc}^*$ will be in the
same ratios as for the $\Sigma_{cc}$ and $\Sigma_{cc}^*$.

The diquark distributions \btobc--\fstar\ must be convolved as in
\fragdist\ with experimentally determined meson fragmentation functions
$D_{\bar Q\to M}(z)$ to obtain fragmentation functions to
$\Sigma_{bc}$, $\Sigma_{bc}^*$ and $\Lambda_{bc}$. These distributions
are then subject to Altarelli-Parisi evolution up to collider energies.
As in the case of $\diq$ production, the quark fragmentation functions
are softened by this evolution, and gluon fragmentation functions are
induced.  The detailed effect of the Altarelli-Parisi equations on the
fragmentation functions to $B_c$ and $B_c^*$ is presented in \ericc,
and the discussion given there applies here as well.  Of course, the
fragmentation functions reduce directly to integrated probabilities as
in \fragtot\ and \fragint, quantities which do not evolve and are more
accessible experimentally.

It is interesting to note that at high-energy colliders
the rates for production of these
doubly heavy baryons are comparable to those for the more familiar
quarkonium systems.  Relating our fragmentation probabilities to those
for $c\to\psi$\ericb\ 
probability for $c\to\Sigma_{cc}, \Sigma_{cc}^*$ to be
$\sim2\times10^{-5}$, for $b\to\Lambda_{bc}$ to be
$\sim2\times10^{-5}$, and for $b\to\Sigma_{bc},\Sigma_{bc}^*$ to be
$\sim 3\times10^{-5}$.  The probabilities for
$b\to\Sigma_{bb},\Sigma_{bb}^*$, $c\to\Lambda_{bc}$ and
$c\to\Sigma_{bc},\Sigma_{bc}^*$ are down by roughly $(m_c/m_b)^3$, or
two orders of magnitude.  Hence it may be possible to observe the
doubly heavy baryons $\Sigma_{cc}^{(*)}$, $\Lambda_{bc}$ and
$\Sigma_{bc}^{(*)}$ at the Tevatron.

There will be additional contributions to the fragmentation to doubly
heavy baryons from radially excited diquark states which subsequently
decay to the ground state.  Equations \fragint\ and \btobc\ also hold
for these processes, using the appropriate value for the diquark
wavefunction at the origin.  Since for the $\psi$ system,
$(R_{\psi(2S)}/R_{\psi})^2\simeq 0.4$, we expect that
$(R_{(cc(2S))}/R_{(cc)})^2$ is also not particularly small and that
excited diquarks will contribute significantly to the production of
doubly heavy baryons.

The largest uncertainty in our calculation arises from our lack of
knowledge of the diquark wavefunction at the origin. We have
na\"\i vely scaled the values for quarkonium systems by a colour factor
of $(1/2)^3$, which is valid only for wavefunctions living entirely in
the Coulombic region of the potential.  This is certainly a poor
approximation for $cc$, $bc$  and $bb$ bound states.
However, we note that in deriving \fragint\ and \btobc\ we
have assumed nothing about the potential except that the diquark
is sufficiently tightly bound to have a $\sim 100\%$ probability of
fragmenting to baryons: $\int_0^1 dy\, D_{(QQ)\rightarrow QQq}(y)
=\int_0^1 dy\, D_{\bar Q\rightarrow M}(y)= 1$.  In reality, this
integrated probability is somewhat less than one, since the diquark may
dissociate as it hadronises, for example $(cc)\rightarrow D D+X$.  Thus
it is likely that \fragint\ and \btobc\ somewhat overestimate the true
fragmentation probabilities.

In summary, we have calculated the fragmentation functions of a heavy
quark $Q$ to a doubly heavy baryon of the form $QQq$ or $QQ'q$. The
perturbative part of the calculation can be related simply to the
fragmentation function to quarkonium, and is of a similar magnitude.
The nonperturbative part may be related to the fragmentation of heavy
quarks to heavy mesons, which may be measured experimentally. We have
used our results to estimate fragmentation probabilities of $b$ and $c$
quarks to doubly heavy baryons of the form $ccq$, $bcq$ and $bbq$.  We
find the production rates for doubly heavy baryons are large enough
that some of these states may eventually be observable.

\bigskip

This work was supported in part by the Department of Energy under
contracts DE--AC03--76SF00515 (SLAC), DE--FG03--90ER40546 (UC San
Diego) and DE--FG03--92ER40701 (Caltech). MJS acknowledges the support
of a Superconducting Supercollider National Fellowship from the Texas
National Research Laboratory Commission under grant  FCFY9219.

\vfil\eject

\listrefs
\vfil\eject
\insertfig{Fig.~1.  Feynman diagrams responsible for the
fragmentation $c\to\diq\cbar $.}{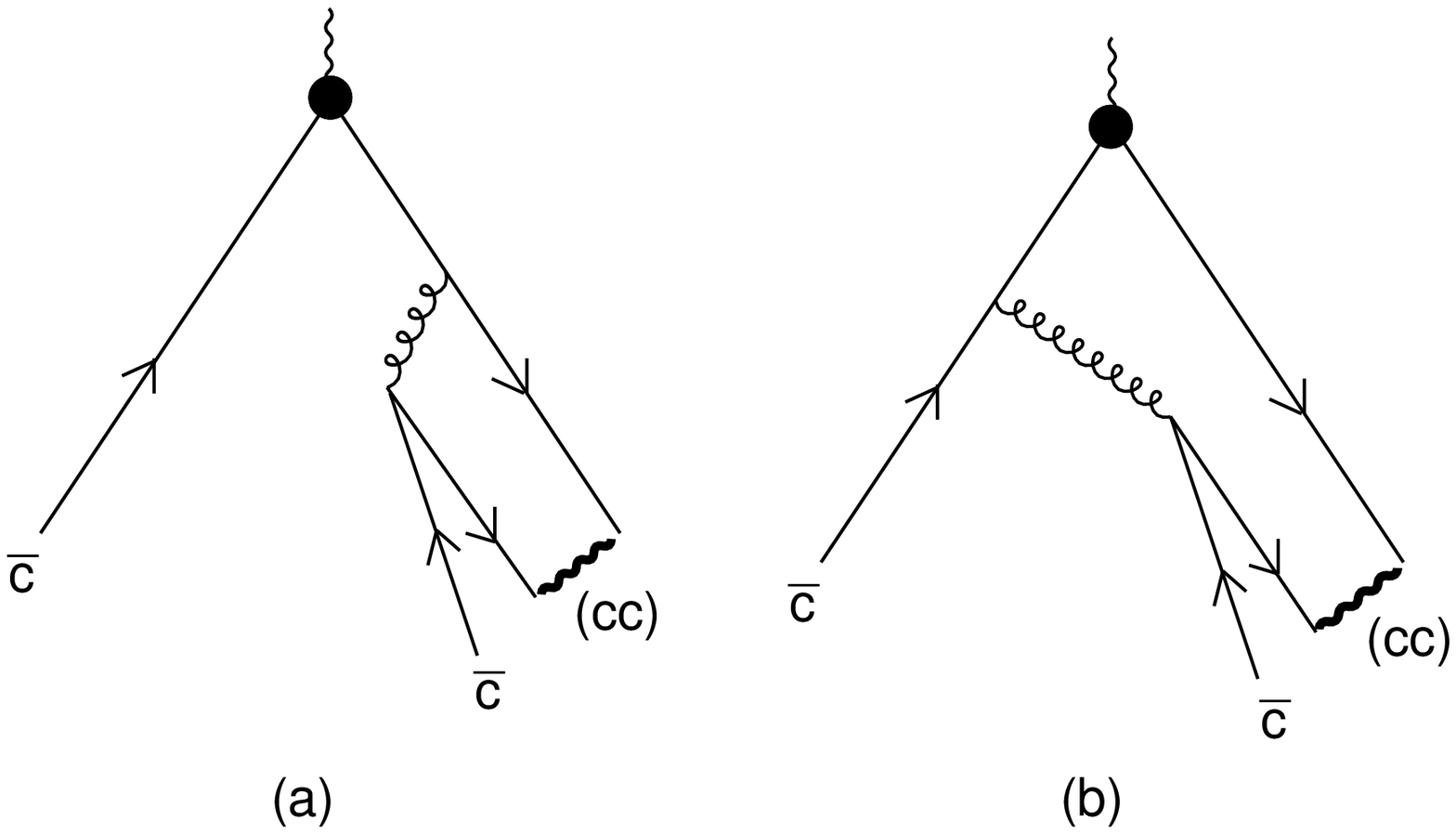}
\vfil\eject
\end